\documentclass[journal=jacsat,manuscript=article, layout=twocolumn]{achemso}
\usepackage{chemformula} 
\usepackage{color,soul}
\usepackage[T1]{fontenc} 
\usepackage{amsmath,amssymb}

\usepackage{tabularx}
\usepackage{array}
\usepackage{dcolumn}
\usepackage{setspace}
\usepackage{url}

\usepackage{float}
\usepackage{caption}
\usepackage{floatflt}
\usepackage{graphicx}

\usepackage[fontsize=11pt]{scrextend}

\author{Dmitrii~S.~Kalashnikov}
\affiliation{Advanced Mesoscience and Nanotechnology Centre, Moscow Institute of Physics and Technology, 141700 Dolgoprudny, Russia}
\alsoaffiliation{National University of Science and Technology MISIS, 119049 Moscow, Russia}

\author{Vsevolod~I.~Ruzhitskiy}
\affiliation{Dukhov Research Institute of Automatics (VNIIA), 127055 Moscow, Russia}
\affiliation{Skobeltsyn Institute of Nuclear Physics, Lomonosov Moscow State University, Moscow, 119991, Russia}
\alsoaffiliation{National University of Science and Technology MISIS, 119049 Moscow, Russia}

\author{Andrey~G.~Shishkin}
\affiliation{Advanced Mesoscience and Nanotechnology Centre, Moscow Institute of Physics and Technology, 141700 Dolgoprudny, Russia}
\alsoaffiliation{Dukhov Research Institute of Automatics (VNIIA), 127055 Moscow, Russia}

\author{Igor~A.~Golovchanskiy}
\affiliation{Advanced Mesoscience and Nanotechnology Centre, Moscow Institute of Physics and Technology, 141700 Dolgoprudny, Russia}

\author{Mikhail~Yu.~Kupriyanov}
\affiliation{Skobeltsyn Institute of Nuclear Physics, Lomonosov Moscow State University, Moscow, 119991, Russia}
\alsoaffiliation{National University of Science and Technology MISIS, 119049 Moscow, Russia}

\author{Igor~I.~Soloviev}
\affiliation{Skobeltsyn Institute of Nuclear Physics, Lomonosov Moscow State University, Moscow, 119991, Russia}
\alsoaffiliation{Dukhov Research Institute of Automatics (VNIIA), 127055 Moscow, Russia}
\alsoaffiliation{National University of Science and Technology MISIS, 119049 Moscow, Russia}

\author{Dimitri~Roditchev}
\affiliation{Laboratoire de Physique et d’Etudes des Materiaux, LPEM, UMR-8213, ESPCI-Paris, PSL, CNRS, Sorbonne University, 75005 Paris, France}

\author{Vasily~S.~Stolyarov}
\affiliation{Advanced Mesoscience and Nanotechnology Centre, Moscow Institute of Physics and Technology, 141700 Dolgoprudny, Russia}\alsoaffiliation{Dukhov Research Institute of Automatics (VNIIA), 127055 Moscow, Russia}
\alsoaffiliation{National University of Science and Technology MISIS, 119049 Moscow, Russia}

\title[An \textsf{achemso} demo]{Josephson vortex-based memory}

\begin{document}

\begin{abstract}
Josephson junctions are currently used as base elements of superconducting logic systems. Long enough junctions subject to magnetic field host quantum phase 2$\pi$-singularities --- Josephson vortices. Here we report the realization of the superconducting memory whose state is encoded by the number of present Josephson vortices. By integrating the junction into a coplanar resonator and by applying a microwave excitation well below the critical current, we were able to control the state of the memory in an energy-efficient and non-destructive manner. The performance of the device is evaluated, and the routes for creating scalable cryogenic memories directly compatible with superconducting microwave technologies are discussed.

\begin{figure}[ht!]
\begin{center}
\includegraphics[width= 8.4 cm]{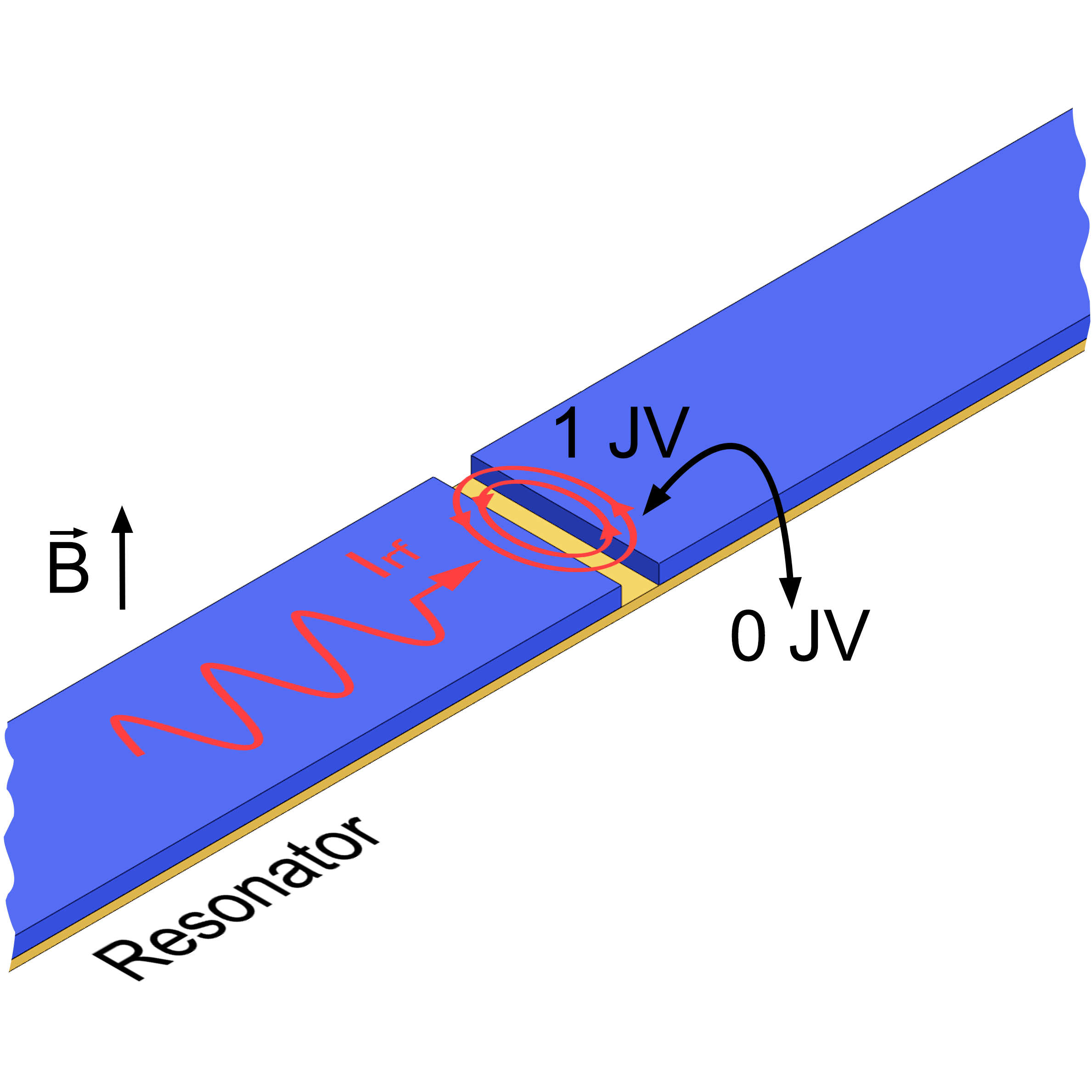}
\end{center}
\end{figure}
\end{abstract}

\section{Introduction}

Superconducting electronics is expected to outperform semiconductor technologies in terms of speed and energy efficiency in a variety of applications, including certain media processing tasks, cryptography, artificial intelligence, digital and quantum computing~\cite{IRDS2022}. Recent efforts in the field led to realization of digital logic \cite{RSFQ_1991,Holmes2013,Beil2017,RSFQ_2021, RSFQ_2023,Tanaka2023}, quantum computers \cite{qubits_review_2010,QRevDS,Soloviev2014,Krantz2019,qubits_review_2020,google,Vozhakov}, neuromorphic systems \cite{neuron_2018,Klenov2018,neuron_2021,neuron_2022, Skryabina2022,Schneider2022,neuron_2023}, among many others. Compatible cryogenic memory is critical for the self-sustainability  of these devices. Current implementations of superconducting memories use superconducting loops involving Josephson junctions (JJs)~\cite{memory_RSFQ_2006, memory_RSFQ_2019, memory_RSFQ_2021, memory_SQUID_2021}, hybrid superconductor/ferromagnet structures \cite{memory_ferrom_2012_Larkin, memory_ferrom_2014, memory_ferrom_2015,Bakurskiy2018,memory_ferrom_2018}, superconductor/ferroelectric elements~\cite{memory_ferroe_2021, memory_ferroe_2022}, and also Abrikosov vortices~\cite{memory_AV_Krasnov} which are magnetic flux quanta inside a superconductor.

In a recent theoretical paper~\cite{JV_memory_Giazotto}, the authors proposed using Josephson vortices (JVs) for the information storage. These vortices appear in a JJ if their length becomes larger than the Josephson penetration depth, $\lambda_J$. The fundamental studies of JVs and, specifically, of their dynamics are numerous; they can be found in several  papers~\cite{JV_Cirillo_1998, JV_Ustinov_1998, JV_Wallraff_2003, JV_Gulevich_2006, JV_Golovchanskiy_2017} and in references therein.
The principle of operation of JV-based memory takes advantage of the hysteretic behavior of a long JJ in an external magnetic field, which was briefly described already in 1965 by Brian Josephson~\cite{Josephson_1965} and studied in detail in several theoretical works~\cite{Josephson_vortex_theory1, Josephson_vortex_theory2}. In fact, similarly to Bean-Livingston barrier for Abrikosov vortex penetration into a superconductor~\cite{bean1964surface,saint1969type,tinkham2004introduction,Schmidt2013physics,likharev1974effect}, there
 exists a barrier for JV entry a long ($\gg \lambda_J$) JJ; these barriers depend on junction geometry, used materials, etc. Their presence resulted, for instance, in the resistance oscillations in a magnetic field observed in layered superconductors containing native JJs~\cite{JV_hysteresis_HTS_2002, JV_hysteresis_HTS_2014}. Thus at the same magnetic field, a long JJ can be found not only in the lowest energy state but also in one of metastable states with different number of JVs inside, depending on the history of the system.

In this paper, we demonstrate the realization of a superconducting memory element based on bi-stability of long JJ. The controlled switching between the zero-JV state of the junction and its one-JV state is obtained by applying pulses of an external magnetic field directed perpendicularly to the device plane. The state of the Josephson junction is read by the microwave current. This operation is possible due the dependence of the JJ impedance on the number of JVs present in inside the junction. The sensitivity of the detection is enhanced by integrating the JJ in the  microwave coplanar resonator. Importantly, the state of the JJ is identified (read) even at currents much lower than its critical current $I_c$. The use of sub-critical currents  has several advantages as compared to read operations requiring the measurements of $I_c$:  enhanced energy efficiency of the memory and non-destruction of its initial state. We analyze the performance of the memory element and discuss the possibilities for improving it based on the results of the developed theoretical model.

\begin{figure*}[ht!]
\begin{center}
\includegraphics[width=17cm]{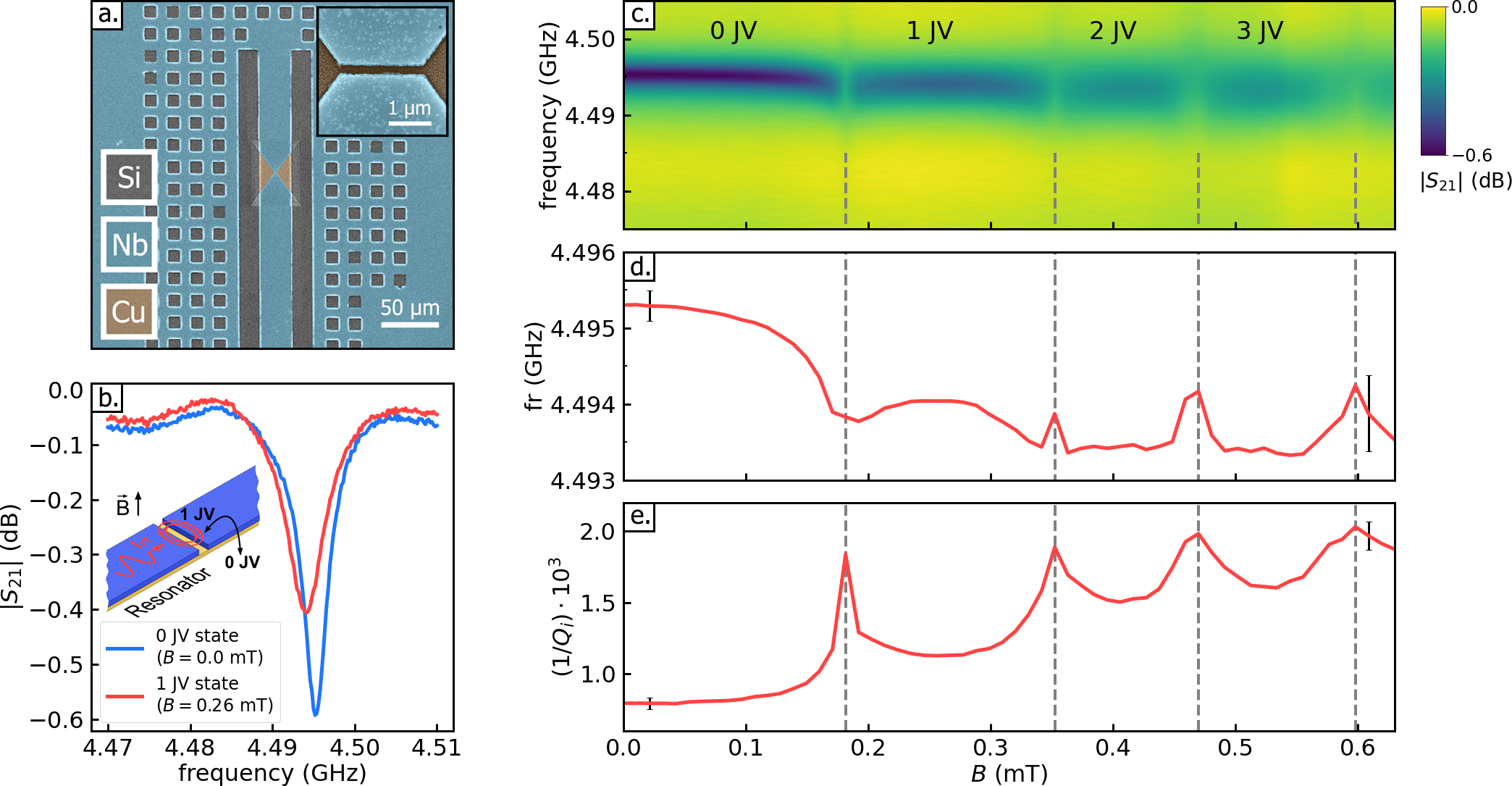}
\caption{Microwave detection of Josephson vortex inside a long SNS junction. a) artificially colored SEM image (top view) of a 4.5 GHz Nb/Cu/Si microwave resonator with a 2 $\mu$m long planar Nb-Cu-Nb junction implemented in the central line (the total length of the resonator is about 7 mm). Inset: a zoom on the junction area; b) resonance modes of the resonator observed as sharp dips in $S_{21}(f)$ signal at zero-applied magnetic field (blue curve) and at $B$=0.26 mT (red curve). Inset: a scheme of microwave measurements of JVs; c) color plot of $S_{21}(f,B)$. Vertical dashed lines delimit regions with different fixed numbers of JVs present in the junction; d) magnetic field dependence of the resonance frequency and, e) of the inverse internal quality factor. The data in d) and e) are extracted from $S_{21}(f,B)$ (see in the text).}
\label{fig: Res3_all}
\end{center}
\end{figure*}

\section{Experiment}

\subsubsection{RF response and the junction state}

In the present work, four identical planar $L=$1.9 $\mu$m long Nb/Cu/Nb 
 SNS Josephson junctions (see inset in Figure~\ref{fig: Res3_all}\,a) were studied. From additional DC current experiments, we determined the value of their critical Josephson current $I_c \approx 940\,\mu$A and estimated the Josephson penetration length to $\lambda_J = 245$\,nm. Consequently, $L \approx 7.6\,\lambda_J$, and thus the junction is long enough to accommodate several Josephson vortices. The junctions were located in the central line of the 1/4-wave coplanar resonators near their ends. An example of the JJ integrated into a 4.5 GHz resonator is presented in Figure~\ref{fig: Res3_all}\,a. The resonators were notch-connected to the common coplanar waveguide used for both, the microwave excitation and the transmission measurement. The resonators have different resonant frequencies enabling addressing them separately.  At a fixed input power $P$ and magnetic field $B$, the frequency $f$ of the microwave signal is swept and the transmitted power $S_{21}(f)$ are registered. It should be emphasized that the measurements were provided at low input power $P = -90$\,dBm, corresponding to AC currents flowing through the junction $\sim$\,4\,$\mu$A, which is by two orders of magnitude lower than $I_c$. Thus, the JJ never reached the normal state. Also, the excitation frequency range, $f = 3-5$\,GHz, was chosen well below the critical (Josephson) frequency $f_c = 2eV_c/h \approx 70$\,GHz, enabling us to consider the JJ as a linear system. 

Figure~\ref{fig: Res3_all}\,b shows two dips in $S_{21}(f)$ measured at zero-field (blue curve) and at 0.26 mT (red curve). They are due to the absorption of the microwave by the resonator.
The curves are Lorentz-shaped and can be described by the resonant frequency, $f_r$, the internal quality factor, $Q_i,$ of the resonator, including the losses in the resonator itself and in the JJ, and the coupling quality factor, $Q_c$, linked to the energy leak from the resonator to the measurement coplanar line. These parameters can be obtained by fitting the experimental data by the expression~\cite{Probst}:

\begin{equation}
    \label{S_21}
    S_{21}(f) = 1 -\frac{Q_l / Q_c}{1 + 2 i Q_l (f / f_r - 1)},
\end{equation}

where $Q_l = (Q_i^{-1}+Q_c^{-1})^{-1}$ is the total quality factor.

Figure~\ref{fig: Res3_all}\,c is a color plot of $S_{21}(f)$ as a function of rising magnetic field 0 $\rightarrow$ 0.65 mT. It demonstrates the existence of 5 field windows characterized by a smooth evolution $S_{21}(f,B)$. Each window is associated with a fixed number, from 0 to 4, of JV inside the junction. The regions are separated by sharp transitions (denoted by vertical dashed lines). Figures~\ref{fig: Res3_all}\,d,e detail the magnetic field evolution of the resonant frequency $f_r(B)$ and of the inverse internal quality factor $Q_i^{-1}(B)$. These parameters are directly related to the complex impedance $Z_L = R_L + i X_L$ of the Josephson junction that adds to the impedance $Z_0 \gg Z_L$ of the resonator itself  and detunes the resonant frequency and the quality factor~\cite{Gao}: 
\begin{equation}
    \label{formula_resonator}
    f_r = f_r^0\left( 1 - \frac{2}{\pi} \frac{X_L}{Z_0} \right); ~~~~ \frac{1}{Q_i} = \frac{1}{Q_i^0} + \frac{4}{\pi} \frac{R_L}{Z_0},
\end{equation}

where $f_r^0$ and $Q_i^0$ are the resonant frequency and the quality factor of the resonator itself, while $R_L$ and $X_L$ represent, respectively, the effective resistance and inductance of the JJ. Table~\ref{table: Impedances} summarizes the essential parameters of the resonator extracted from the data for different numbers $N$ of JVs present.
\begin{table*}[ht!]
\caption{Resonance frequency $f_r$, inverse internal quality factor $1/Q_i$ and variations of the real $R_L$ and imaginary $X_L$ parts of the complex impedance $Z_L$ for different numbers $N$ of JVs in the JJ extracted from data in Fig.\,\ref{fig: Res3_all}c using the formulas~\ref{S_21} and~\ref{formula_resonator}}
\begin{center}
\begin{tabular}{c c c c c}
JJ state, N   & $f_r$, GHz    & $1/Q_i \cdot 10^{3}$   & $R_L(N)-R_L(0)$, m$\Omega$   & $X_L(N)-X_L(0)$, m$\Omega$   \\ \hline
0 JV       & 4.4953       & 0.79    & 0                         & 0                         \\ 
1 JV      & 4.4940       & 1.13    & 13                        & 22                        \\ 
2 JV      & 4.4934       & 1.50    & 28                        & 33                        \\ 
3 JV      & 4.4934       & 1.60    & 32                        & 33                        \\ 
\end{tabular}
\end{center}
\label{table: Impedances}
\end{table*}
It becomes immediately clear that the $R_L(N)$ increases with $N$. Obtained dependence qualitatively coincides with the calculated curve in the work \cite{Impedance_JJ} where the surface impedance of the infinite Josephson contact in the external magnetic field was investigated. The reason for this additional dissipation is the forced motion of JVs inside the JJ under microwave excitation. Indeed, since $f \ll f_c$, and the amplitude of microwave current is well below $I_c$, the excitation can be seen as a source of a tiny oscillatory Lorentz force that slightly shakes the JV around its equilibrium position inside the junction. Note that in ideal SIS JJs the JV motion is dissipation-less; however it is not so in SNS JJs because of quasiparticles in the N-region. The minimum of dissipation expectantly occurs when no JV are present. At $N=1$, the dissipation parameter $R_L$ increases by 13 m$\Omega$; the fact that this is by an order of magnitude lower than the normal JJ resistance $R_N = 150$ m$\Omega$ is consistent with the expected low amplitude of JV forced motion due to excitation. At $N=2$, increases by 28 m$\Omega$. This doubling of dissipation is simply explained by two JVs moving back and forth along the junctions, instead of one. However, adding one more JV does not triple the dissipation. A possible reason could be related to the limited size of the junction, $L \approx 7.6\,\lambda_J$, that can only accept 3-4 JVs at maximum. As $N$ increases, JVs start to interact, reducing the effect of microwave excitation per vortex. Note also, that for $N>2$ the precision of determining $R_L$ and $X_L$ drops significantly and cannot be considered as reliable anymore.

We can now understand the global oscillatory evolution of $Q_i^{-1}(B)$ presented in Figure~\ref{fig: Res3_all}\,e. For each $N$, there exists a filed value corresponding to the minimum energy and to the most stable equilibrium for JVs in the junction; it is zero for $N$=0, 0.26 mT for $N$=1, 0.41 mT for $N$=2, etc. At these fields, the amplitude of JV forced motion and the related dissipation $R_L$ are lowest, and $Q_i^{-1}(B)$ experiences local minima. When the magnetic field deviates from these values, the potential in which evolve JVs flattens; this results in an increase of the forced motion amplitude and, consequently, of the dissipation; $Q_i^{-1}$ raises. Close to the field values corresponding to the transitions $N \rightarrow N\pm 1$ (denoted by vertical dashed lines in Figure~\ref{fig: Res3_all}) the potential flattens, JV motion amplitude is the highest, the dissipation is strongest, resulting in peaks in $Q_i^{-1}$. It has to be noted that a similar behaviour of JVs in long JJ was observed in works \cite{Dremov_2019, Grbenchuk_2020, Hovhannisyan_2021, Stolyarov_2022} where the JJ was coupled to a mechanic resonator represented by the oscillating magnetic cantilever of the MFM microscope.  Similarly to microwave excitation, the cantilever excites the motion of the JV inside the junction and probes the corresponding energy dissipation. The authors showed that the highest energy dissipation occurs when individual JVs enter or exit the junction.


\subsubsection{Vortex hysteresis effect}

\begin{figure}[ht!]
\begin{center}
\includegraphics[width=8.4cm]{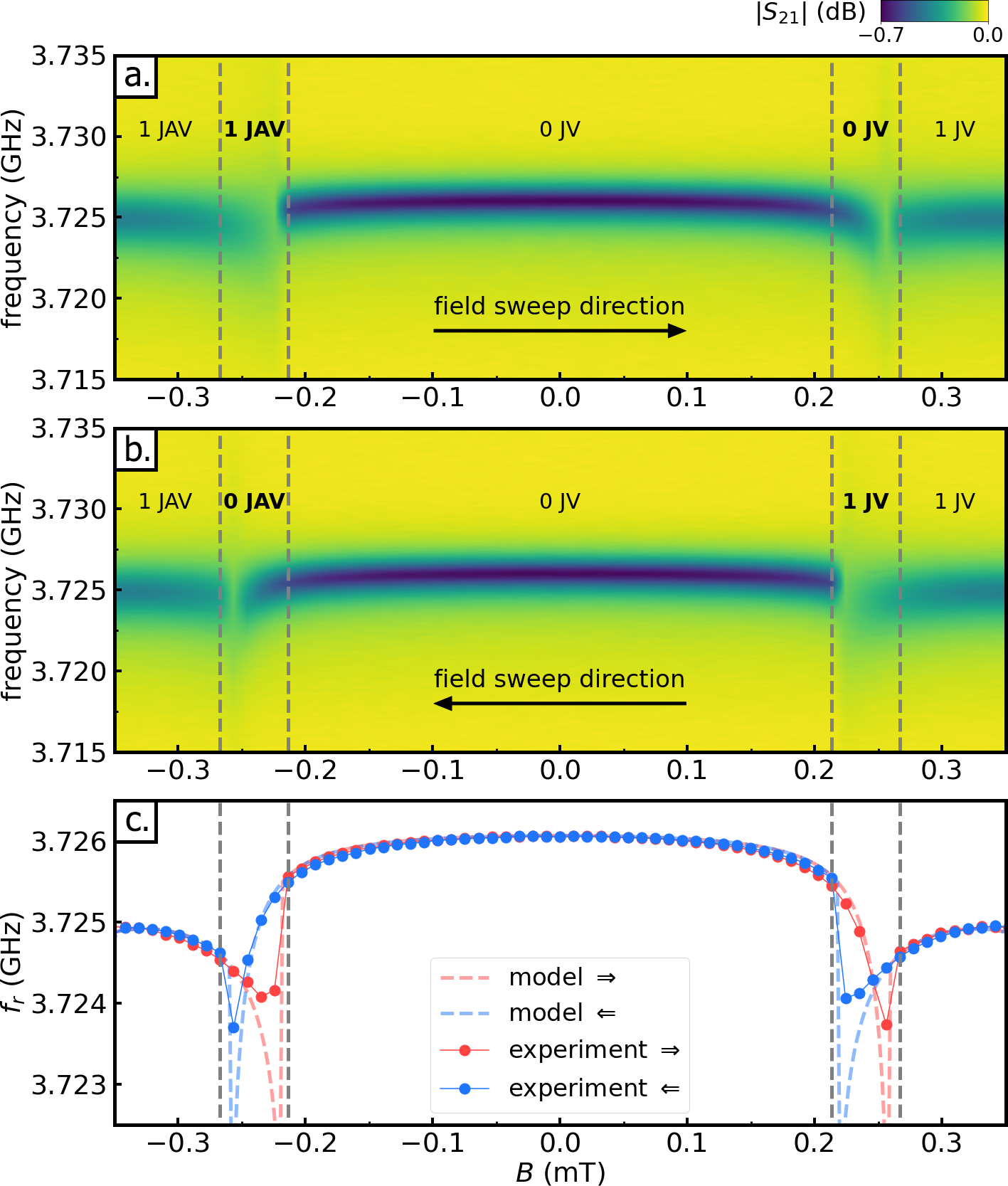}
\caption{Hysteresis in $N \rightleftarrows N+1$ JV transitions (3.73 GHz resonator). Images a) and b) are color plots of $S_{21}(f,B)$ recorded at opposite field sweep directions (marked by arrows). The hysteresis regions are delimited by vertical dashed lines; c) variations of the resonant frequency as the field is swept from negative to positive values (red dots) and back (blue dots). Color dashed lines are corresponding numerical fits (further explanations in the text).}
\label{fig: Hysteresis_res1}
\end{center}
\end{figure}

For the implementation of a memory element, the key effect is the hysteresis of the system in a magnetic field near $N \rightarrow N\pm 1$. 
Figures~\ref{fig: Hysteresis_res1}\,a,b show the magnetic field dependencies of $S_{21}(f)$ measured for the opposite directions of the magnetic field sweeps around zero. The obtained patterns are globally symmetric and coincide, except for the two regions, -\,0.27\,mT\,$<B<$ -\,0.21\,mT and +\,0.21\,mT\,$<B<$\,+\,0.27\,mT, where $0 \rightleftarrows \pm 1$ transitions occur. In these regions, both the transition field and $S_{21}(f,B)$ evolution depend on the field sweep direction. Precisely, the resonance frequency lowers smoothly when the field approaches the transition field from the current state but jumps abruptly to the next state. This dynamics always coincides with the direction of the field sweep and therefore it is not a simple displacement of the pattern in a magnetic field. This eliminates the possibility that the observed hysteresis is created elsewhere, by the solenoid, for instance, and not in the Josephson junction. Despite the fact that the hysteresis region is rather narrow, it is perfectly reproducible.

\begin{figure*}[ht!]
\begin{center}
\includegraphics[width=17cm]{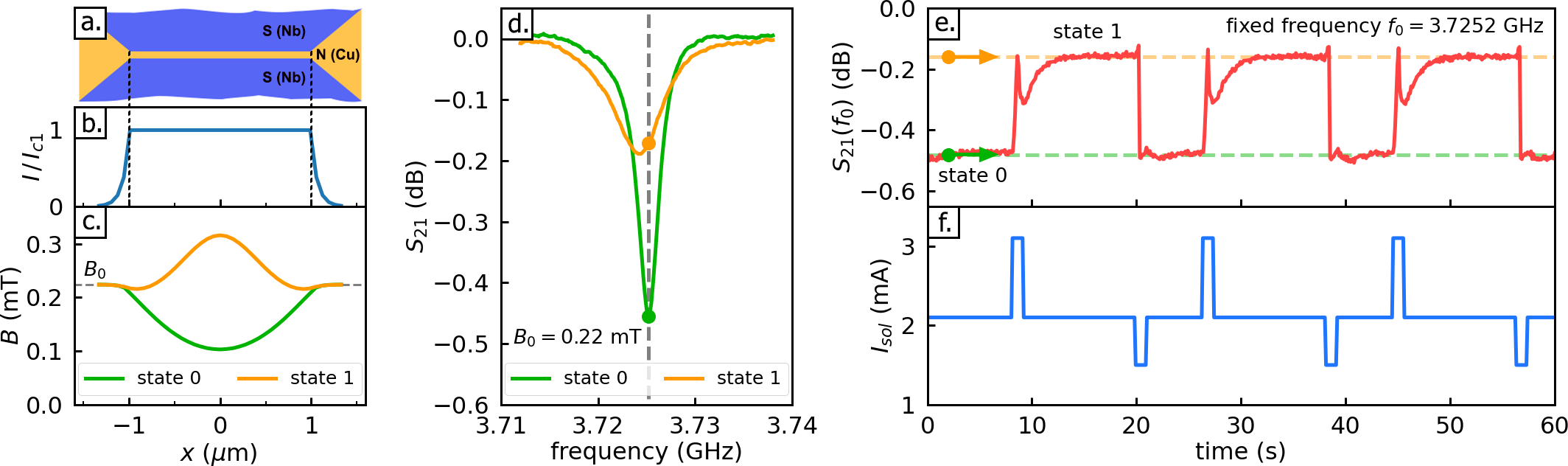}
\caption{
Switching between $N$=0 and $N$=1 JV states. a) theoretically considered geometry of the sample; b) considered distribution of the critical current density along the junction; c) calculated distribution of the magnetic field inside the JJ in the two states at the same external magnetic field  $B_0 = 0.22$\,mT inside the hysteresis region; d) experimentally measured resonance peaks at the same field; e) experimental demonstration of the switching detection in $S_{21}$ measured at a fixed frequency $f_0 = 3.7252$\,GHz; f) a series of current pulses applied to the solenoid to vary the magnetic field and switch between the two states (plots e) and f) share the temporal axis).}
\label{fig: States switching}
\end{center}
\end{figure*}

Our numerical simulations of the system's behaviour robustly supported the experimental data and working hypotheses and enabled a deeper insight in the microscopic processes inside the junction. The details about the numerical model can be found in the Methods section. We considered a planar JJ of a geometry reproducing  the experimentally studied one, as presented in Figure\,\ref{fig: States switching}\,a. The critical current distribution in the junction was taken constant inside the junction and exponentially decaying outside, Figure\,\ref{fig: States switching}\,b. Using such simple parameters, the model successfully reproduced the entry and exit of JVs, the hysteresis effect and the observed transition asymmetry in $f_r(B)$. Precisely, the model curves (blue and red dashed lines in Figure\,\ref{fig: Hysteresis_res1}\,c) nicely follow the experimental data points and reproduce a smooth decrease in the resonant frequency when exiting the current state, and then a sharp dynamics when entering the following state. Though in the model, the resonant frequency drop was found deeper than in the experiment, down to $f = 3.721$\,GHz. This occurs because the model does not take into account the losses in the resonator, a non-uniform focusing of the magnetic field due to the flat geometry or a JV shaking by microwave excitation.
Figure\,\ref{fig: States switching}\,c shows the distribution of the magnetic field inside the Josephson junction obtained from the model in an external magnetic field equal to $B_0 = 0.22$\,mT, which corresponds to the hysteresis region. This result shows that at the same field, the JJ can indeed be in two different stable states, characterized by the number (0 or 1) of JVs inside. The local field distribution inside JJ is different: in $N$=0 state, the external field is partially screened towards the junction center by Meissner currents (green line), while in $N$=1 state the JV currents increase it (yellow line). Figure\,\ref{fig: States switching}\,d shows the experimentally measured resonance curves corresponding to these two states, measured at the same field 0.22 mT. The resonant frequencies and quality factors clearly differ enabling a reliable distinguishing between them.

\subsubsection{Switching between JV states}

To demonstrate a possible working principle of the JV-based memory cell, we performed an experiment in which the switching between $N$=0 and $N$=1 states was driven by magnetic field. The results are presented in Figures\,\ref{fig: States switching}\,e,f. The current through the superconducting solenoid was first set to zero and then to the value corresponding to the field $B_0$=0.22 mT. At this ``write zero'' condition, the system is in $N$=0 state. The signal $S_{21} \approx$ -0.5 dB corresponds to a non-destructive ``read'' of ``zero'' state of the memory. Then, an additional current pulse of $\sim$1 sec duration was applied to temporarily increase the field and drive the JJ to $N$=1 state. A single Josephson vortex penetrates the JJ and remains there when the field returns to $B_0$; this is ``write one'' operation. The signal $S_{21}$ raises to $\approx$ -0.2 dB; that corresponds to a non-destructive ``read'' of ``one'' state of the memory. Similarly, when negative current pulse is applied, the field decreases below the hysteresis region, the Josephson vortex exits the junction, and when the field returns back to the value $B_0$ the JJ remains in $N$=0 state --- ``zero'' state of the memory. The write/read processes are perfectly reproducible. This demonstrates the system to work indeed as a memory element. Note that in the experiment the used superconducting solenoid has a long relaxation time $\tau = {L/R} \sim 10^0$\,s, due to a large inductance $L$ of the solenoid and a low resistance $R$ of the in-parallel connected ``open'' superconducting switch. That explains why the detection signal $S_{21}$ does not change immediately but evolves in a seconds. 
Our model showed that in practice, the pulses required for successful write processes can be as short as $\sim$ 40\,ps, which enables a state-of-the art fast ``write'' operations \cite{IRDS2022}.

\subsubsection{Towards competitive JV-based memories}

The main goals of the present work were to suggest a new kind of cryogenic memory based on a single Josephson vortex in SNS junctions, dress the microscopic picture of the JV dynamics and to demonstrate one of possible working principles of JV-based memory function. Nevertheless, the obtained experimental results already enable us to foresee several advantages of these memories and envisage some routes for the performance optimisation.

Currently, cryogenic microwave devices integrating JJ are actively developed, such as generators \cite{laser_Science_2017, generator_NE_2021}, amplifiers\cite{PARAMP_2015, PARAMP_2020}, tunable resonators \cite{Tunable_res_2008, Tunable_res_2019, Tunable_res_2023} or impedances \cite{Habib}. With this respect, the use of a similar technology and signal transmission method in the present work is advantageous as it makes our JV-based memories on-chip compatible with the existing rich family of superconducting microwave devices. Furthermore, the results of the present work open promising options for the development of advantageous quantum-classical digital/analogue interfaces for quantum computers. Indeed, the transmission of the microwave signals that control the quantum core, with its simultaneous noise isolation, is a bottleneck that is currently seen as the main challenge to further progress in increasing the number of qubits in quantum processors. The classical control system should be placed in a fridge nearby the quantum core, and possess high speed and energy efficiency of its digital part \cite{McDermott2018}, while being easy compatible with microwave part of electronics communicating with qubits. The proposed memory is exactly what is desired, since it can directly affect the microwave characteristics of analogue devices while being controlled by digital single-flux-quantum (SFQ) circuits \cite{RSFQ_1991,IRDS2022}, where the information is presented as JVs. This compatibility can be also highly sought after in hybrid digital/analogue implementations of superconducting neural networks \cite{Islam2023}.




The ability to encode several JV states at once is also advantageous as it makes possible to envisage other devices such as multi-bit memories, registers or counters \cite{Fujiwara2002,Katam2020,Jardin2023}. Also, the use of artificial uniformity suggested in Figure\ref{fig: States switching}\,a enables a spatial localization of the vortex position. In principle, this could allow one to access the state of a specific bit. In addition, already in the studied simple linear JJ geometry, anti-vortices (JVs of the opposite flux direction) can also be used, thus forming basic elements of ternary logic, with '-1', '0' and '1' encoded states.



The speed of write/read operations and related energy dissipation are important parameters of cryogenic memories. In the experiment, the ``write'' pulse duration was very long, about $\sim$ 1 s, because of a very large solenoid used. Though, in realistic memory elements, the JJ could be driven locally by a tiny current loop; the pulse duration could quite easily be reduced to $\sim$ 10$^{-9}$ s, at least. The energy dissipated in the JJ upon ``write'' process is that required to put (remove) a single JV in (from) the JJ; it depends on JJ characteristics and can be estimated in the framework of our numerical model; for the studied JJs one gets $\sim$ 1 aJ. The ``read'' process at resonance requires a time $\sim Q_l\times T$, where $T$ is the period of the microwave excitation. For the studied (non-optimized) device one gets $\sim 10^{-7}$ s, which is quite long. Though, ``read'' operation is much less dissipative than ``write'' one, since the shaking microwave current and $R_L$ are well below $I_c$ and $R_N$. As a result, the total energy dissipated in the JJ during one ``read'' operation remains far below 1 aJ.

Evidently, at this early stage of development, the studied device was not optimized for serving as a JV-based memory element. Hereafter we list some ideas for improvements in the future. First, the write time can be reduced. Clearly, using large solenoids is not appropriate. Instead, tiny local superconducting loops could be designed for JV manipulation. Also, increasing the critical voltage $V_c$ of the junction could be interesting. The speed of processes in SNS is determined by the critical frequency $f_c$, which is proportional to $V_c$. Our sample was far from the maximum possible value, that can reach several mV \cite{SINIS_2021}. In this way, the writing times can be reduced to about 6~ps.

Second, the difference in the detection signal $S_{21}$ corresponding to the two states can be increased enabling their easier identification. This can be done by increasing the absorption of the resonator. According to the expression~\eqref{S_21}, at resonance frequency $f=f_r$, the transmission coefficient is equal to $S_{21}(f_r) = 1 - Q_c/Q_l = 1/(1+Q_i/Q_c)$. That is, an increase in the $Q_i/Q_c$ ratio can lead to almost complete signal absorption. For example, if our system had the internal Q-factor $Q_i$=50000, the difference in output power would be 84\%. Note that the values of $Q_i$ and $Q_c$ can be greatly varied in microwave coplanar resonators and reach values of the order of $10^6$ \cite{Q-high_res_2021}. However, large Q-factors would lead to long relaxations of the signal in the resonator. Therefore, the optimisation is necessary, depending on a specific use of the memory.

\begin{figure}[ht!]
\begin{center}
\includegraphics[width=8.4cm]{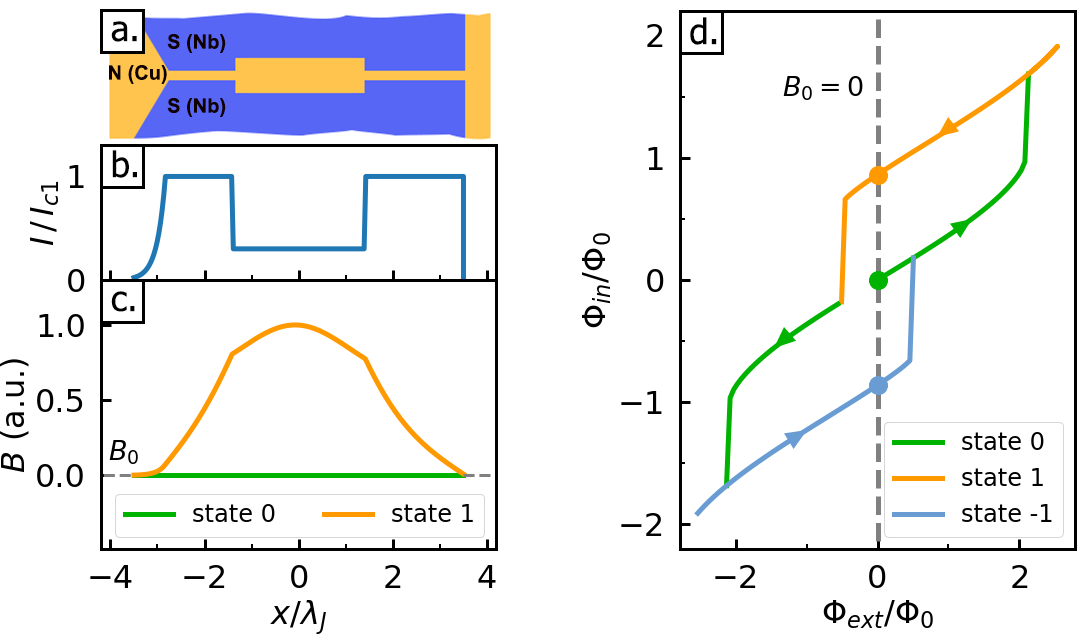}
\caption{
Suggestion for the memory element working at zero-magnetic field. a) top view of the junction; b) the width modulation leads to a non-uniform distribution of the critical current density; c) distribution of the magnetic field inside the junction at zero-external field for $N$=0 state (green line) and $N$=1 state (yellow line); d) dependence of the total magnetic flux through the junction as a function of the external magnetic flux is hysteretic. Due to the peculiar geometry of the JJ, the hysteresis region 0~$\rightleftarrows \pm$1 includes $\Phi_{ext}$ = 0 (vertical dashed line).
}
\label{fig: Improvement}
\end{center}
\end{figure}

Third, it would be interesting to avoid applying a permanent non-zero working field $B_0$. This cannot be achieved in standard long JJs because, according to the works\cite{Josephson_vortex_theory1, Josephson_vortex_theory2}, in JJs of a finite length there are no quasi-equilibrium states at a zero-field. For $0\rightleftarrows1$ hysteresis to occur around zero-field, it is necessary to artificially strengthen the barriers for the JV entry. This can be done, for example, by increasing the density of the critical current at the edges of the junction. Figure~\ref{fig: Improvement}\,a suggests a sketch of such a junction which is narrower at its ends and wider in the center. Numerical simulations of such a junction take into account a non-uniform critical current density distribution, Figure~\ref{fig: Improvement}\,b. They show that such a JJ indeed enables +1 and -1 quasi-equilibrium JV states at a zero-field along with the true equilibrium state 0 with no magnetic field inside the junction. The local field distribution in these states is shown in Fig~\ref{fig: Improvement}\,c . The full hysteresis curve of such a junction upon external field cycling is shown in Fig.~\ref{fig: Improvement}\,d. Now, the working field can be selected equal to zero (vertical dashed line). In addition, it would be interesting to explore the suggestion \cite{Roditchev_2015} of generating JVs states by purely electric means.









\section{Conclusion}

To conclude, we presented a new kind of cryogenic memory in which the information is encoded in individual Josephson vortex present in a long Superconductor - Normal metal - Superconductor Josephson junction. The junction is embedded into the superconducting coplanar resonator and thus is directly compatible with the state of the art cryogenic microwave technology. The desired number of vortices in the junction is set by applying calibrated pulses of magnetic field; an energy-efficient non-destructive readout is done by slightly shaking the introduced vortices with a weak microwave excitation (two orders of magnitude below the critical current), and by measuring the microwave response of the resonator. The process is similar to the matchbox shaking and sound listening -- a common way to know if there are matches inside with no need for box opening. We make the first assessment of the memory performance and suggest different routes for its improvement. While for now only 0 and 1 vortex states were exploited to demonstrate a standard memory function, it is straightforward to create multi-state memories and more complex logic elements by making longer Josephson junctions, by tuning material's properties and shaping the device geometry.


\section{Methods}

\subsection{Device fabrication}

The device consists of four identical Nb/Cu Nb junctions integrated into coplanar microwave resonators of different resonant frequencies. The resonators are notch-coupled to the common coplanar waveguide. The lateral size of the whole device is $\approx$ 9 mm. To realize it, Nb/Cu films were first deposited onto a highly resistive silicon wafer by successive magnetron sputtering of 50 nm copper (bottom layer) and 100 nm niobium (top layer). Then, coplanar structures were fabricated using etching through the resist mask. Niobium was removed by Reactive Ion Etching in CF${}_4$+O${}_2$ plasma; copper --- using wet chemical solution FeCl${}_3$. The superconducting critical temperature of Nb/Cu bilayer, $T = 8.25$\,K, was measured by four-terminal electron transport method.

\subsection{Experimental setup}
In the experiments, the resonators were excited and probed individually at their resonant frequencies (3-5 GHz) using two microwave lines connected to the device. The input line contained several cryogenic attenuators with a total attenuation of -60 dB. The output line was equipped with a cryogenic amplifier with a gain of +40 dB. The parameter $S_{21}$ was measured using a vector network analyzer (VNA). The measurements were provided in a dilution refrigerator at a base temperature of 35 mK. A superconducting solenoid was used to create an external magnetic field, with a field to current conversion constant $B/I_{sol} = 0.1068$~T/A.

\subsection{Numerical model}
A distributed model of a long contact was used similarly to the work \cite{Stolyarov_2022} (see supporting information for more details). The gate-like distribution of the critical current was considered, as shown in Figure \ref{fig: States switching}(b). For the calculations, the junction was divided in 30 segments in its central part, with the same critical current $I_{c1}$ = 32.49 $\mu A$ in each segment; the JJ edges were divided in 5 segments each, with exponentially decaying critical current. The normalized JJ length was taken $L/\lambda_J^{fit} = 4.09$.
For the complex impedance calculations, a weak harmonic excitation current was used, $I/I_{c1} = 0.01\sin(0.01\tau)$. As for the voltage across the junction, we took the half-sum of the derivatives of the superconducting phases at the edges of the junction, $V/V_c = 0.5 (\dot{\varphi_1} + \dot{\varphi_{30}})$.

\section{Acknowledgement}
The authors are grateful to  O.V. Skryabina for useful discussions of the obtained results and assistance in the experimental studies.

\subsection{Funding}

The sample preparation process was supported by the Ministry of Science and Higher Education of the Russian Federation (No. FSMG-2023-0014). The experiments were carried out with the support of the Russian Science Foundation, Grant No. 23-72-30004 https://rscf.ru/project/23-72-30004/. The theoretical modeling of high-frequency spectroscopy of the Josephson vortex dynamics 
and 
switching effects was supported by the Russian Science Foundation,
Grant No. 20-12-00130 (\url{https://rscf.ru/project/20-12-00130/} ).
The concept of cryogenic memory was developed
 with the financial support of the Strategic Academic Leadership Programme “Priority-2030” (NUST MISIS Grant No. K2-2022-029) and French project CrysTop (ANR-19-CE30-0034).

\newpage

\bibliography{Sources}

\end{document}